\documentclass{aa}
\usepackage{epsf}

\begin{document}

 \thesaurus{12          
              (08.14.1;   
               02.04.1)   
            }


\title{Neutrino emission due to Cooper pairing of nucleons
             in cooling neutron stars}
\author{D.G.~Yakovlev\thanks{
E-mail: yak@astro.ioffe.rssi.ru} 
         \and
        A.D.~Kaminker \and  K.P.~Levenfish}
\institute{
        Ioffe Physical Technical Institute,
        Politekhnicheskaya 26, 194021 St.-Petersburg,
        Russia
          }
\date{Received 18 September 1998 / Accepted 29 October 1998}
\offprints{D.G.~Yakovlev}
\titlerunning{ Neutrino emission
        due to Cooper pairing \ldots}
\authorrunning{D.G.~Yakovlev et al.}
\maketitle


\begin{abstract}
The neutrino energy emission rate due to formation
of Cooper pairs of neutrons and protons in the superfluid
cores of neutron stars is studied.
The cases of singlet-state
pairing with isotropic superfluid gap and triplet-state
pairing with anisotropic gap are analysed.
The neutrino emission due to the singlet-state pairing of protons
is found to be greatly suppressed with respect to the
cases of singlet- or triplet-state pairings of neutrons.
The neutrino emission due to pairing of
neutrons is shown to be very important
in the superfluid neutron--star
cores with the standard neutrino luminosity and with the luminosity
enhanced by the direct Urca process.
It can greatly accelerate both, standard and enhanced, cooling
of neutron stars with superfluid cores.
This enables one to interpret
the data on surface temperatures of
six neutron stars, obtained by fitting the observed
spectra with the
hydrogen atmosphere models,
by the standard cooling with moderate nucleon superfluidity.
\end{abstract}

%
\section{Introduction}
Neutron star interiors contain matter
of nuclear and supranuclear density.
Various microscopic theoretical
models predict different compositions of this matter
(neutrons, protons and electrons; hyperons;
kaon or pion condensates; quarks), different
equations of state (from very soft to very stiff)
and superfluid properties of strongly interacting
baryonic components (nucleons, hyperons, quarks).

Neutron stars are born very hot in supernova explosions
but cool gradually in time. Their cooling depends
on properties of stellar matter. Comparison
of theoretical cooling models with observational data
on the surface temperatures of isolated neutron stars
yields a potentially powerful method to constrain
models of superdense matter.

Young and middle-age ($t \la 10^4$--$10^5$ yr) neutron
stars cool mainly via neutrino emission from their
interiors. It is important, thus, to know the main
neutrino production mechanisms. For simplicity, we
restrict ourselves by consideration of neutron
stars whose cores are composed of neutrons ($n$),
with some admixture of protons ($p$) and electrons ($e$).
The neutrino emission mechanisms
in the stellar cores may be divided into two groups,
leading to {\it standard} or {\it rapid} cooling
(e.g., Pethick  \cite{p92}).

The standard cooling lowers the surface temperature
to about $10^6$ K in $t \sim 10^4$ yr.
It goes mainly via the modified Urca process
(e.g., Friman \& Maxwell \cite{fm79}, Yakovlev \& Levenfish \cite{yl95})
\begin{equation}
   n + N \to p + N + e + \bar{\nu}_e,~~~
   p + N + e \to n + N + \nu_e,
\label{Murca}
\end{equation}
and the nucleon--nucleon bremsstrahlung
\begin{equation}
   N + N \to N + N + \nu + \bar{\nu},
\label{Nbrems}
\end{equation}
where $N$ is a nucleon ($n$ or $p$).

Rapid cooling is strongly enhanced by the direct Urca reaction
\begin{equation}
   n \to p + e + \bar{\nu}_e,~~~
   p + e \to n + \nu_e,
\label{Durca}
\end{equation}
which can operate (Lattimer et al.\ \cite{lpph91})
only in the central regions of
rather massive neutron stars with not too soft
equations of state. If the direct Urca reaction
is allowed, the neutrino emissivity is typically 4 -- 5
orders of magnitude higher than in the standard
reactions (\ref{Murca}) and (\ref{Nbrems}), and
the surface temperature decreases
to several times of $10^5$ K in $t \sim 10^4$ yr.

Cooling of neutron stars can be strongly affected
by superfluidity of neutrons and protons in the stellar
cores. The superfluidity is generally thought to be
of BCS type produced under nuclear attraction of nucleons.
At subnuclear densities $\rho \la \rho_0$
(where $\rho_0=2.8 \times 10^{14}$ g cm$^{-3}$ is the standard
nuclear--matter density) the neutron pairing occurs
due to $nn$ attraction in the $^1$S$_0$ state.
The superfluid gaps depend sensitively on
$nn$ interaction model. Various microscopic theories
(e.g., Tamagaki \cite{t70}, Amundsen \&  {\O}stgaard \cite{ao85},
Baldo et al.\ \cite{bcll92}, Takatsuka \& Tamagaki \cite{tt93, tt97})
predict these gaps to vary in the range from some ten keV to
some MeV. However, the singlet-state interaction of
neutrons becomes repulsive at $\rho \sim \rho_0$,
and, therefore, the singlet-state neutron superfluidity vanishes
near the boundary between the
neutron star core and the crust.
Deeper in the core ($\rho \ga \rho_0$),
the triplet-state ($^3$P$_2$) $nn$
interaction can be attractive to produce the
superfluid with an anisotropic gap.
Since the number density of protons is much smaller
than that of neutrons, the singlet-state $pp$ interaction
is thought to be attractive in the stellar core
leading to proton superfluidity.
The superfluidity of $n$ and $p$ affects the main
neutrino generation mechanisms (\ref{Murca}) -- (\ref{Durca})
in the neutron star cores, and hence cooling of
neutron stars with superfluid interiors.
Superfluidity always suppresses
these reactions decreasing the neutrino luminosity
of neutron stars.

However, the appearance of neutron or proton superfluid
in a cooling neutron star initiates
an additional specific neutrino production mechanism
associated with the direct interband transition of a nucleon,
\begin{equation}
      N \to N + \nu + \overline{\nu}.
\label{CPF}
\end{equation}
The mechanism is allowed due to existence of a superfluid gap
in the nucleon dispersion relation. The process may be called
the neutrino emission due to {\it Cooper pair formation}.
It was proposed and calculated for singlet-state neutron
superfluidity
by Flowers et al.\ (\cite{frs76}).
It was rediscovered later
by Voskresensky \& Senatorov (\cite{vs86, vs87}).
Until recently, the process has been `forgotten',
and has not been used in the studies of the neutron star cooling.
It has been included into recent cooling simulations
by Schaab et al.\ (\cite{svsww97}), 
Page (\cite{p98}), Yakovlev et al.\ (\cite{ykl98}) and
Levenfish et al.\ (\cite{lsy98}) although its effect has not
been analysed in details.

In Sect.\ 2 we present the derivation of the neutrino energy
generation rate due to singlet and triplet Cooper pairing of nucleons.
In the particular case of singlet--state $nn$ pairing
we reproduce the result by Flowers et al.\ (\cite{frs76}). The cases of
triplet pairing of neutrons and singlet paring of protons
appear to be different.
In Sect.\ 3 we compare the Cooper-pair neutrino emission
with other neutrino processes in the neutron star cores,
and in Sect.\ 4 we analyse the effect of the Cooper-pair
neutrinos on cooling of neutron stars with superfluid cores.

\section{Cooper-pair neutrino emissivity}

Consider neutrino emission 
due to Cooper pair
formation (\ref{CPF})
of nucleons. In the absence of superfluidity
the process is strictly forbidden by energy--momentum conservation:
a neutrino pair cannot be emitted by a free nucleon.
Superfluidity introduces the energy gap into the
nucleon dispersion relation near the Fermi surface
which opens the reaction.

\subsection{General formalism}

Following Flowers et al.\ (\cite{frs76})
we will study the process (\ref{CPF}) as annihilation
$\tilde{N} + \tilde{N} \to \nu + \bar{\nu}$
of quasinucleons $\tilde{N}$ in a Fermi liquid with creation of
a neutrino pair.
The process goes via electroweak neutral currents;
neutrinos of any flavors can be emitted.
We will assume that nucleons are nonrelativistic and degenerate,
and we will use the approximation of massless neutrinos.
Let us derive the neutrino energy generation rate (emissivity)
due to Cooper pairing of protons or neutrons in a triplet
or singlet state. In this way we will generalize the
results by Flowers et al.\ (\cite{frs76})
to the case of Cooper pairing of protons,
and to the most important case of triplet pairing.
The interaction Hamiltonian ($\hbar = c = 1$) is given
by (e.g., Friman \& Maxwell \cite{fm79})
\begin{equation}
     H_w = - { G_{\rm F} \over 2 \sqrt{2} } \,
     \left( c_V \, J_0 l_0 - c_A \, {\vec J}\cdot{\vec l} \right),
\label{Hw}
\end{equation}
where $G_{\rm F}$ is the Fermi weak-interaction constant, and the terms
containing $c_V$ and $c_A$ describe contributions of the vector
and axial vector currents, respectively.
The factors $c_V$ and $c_A$ are determined by quark composition
of nucleons (e.g., Okun' \cite{o90}), 
and they are different for $n$ and $p$.
For the reactions with neutrons, one has
$c_V=1$ and $c_A=g_A$, while for those with protons,
$c_V=4 \, \sin^2 \Theta_{\rm W} -1 \approx -0.08$ and
$c_A= - g_A$, where $g_A \approx 1.26$ is the axial--vector
constant and
$\Theta_{\rm W}$ is the Weinberg angle ($\sin^2 \Theta_{\rm W} = 0.23$).
Notice that similar interaction Hamiltonian 
describes the neutrino emission due to pairing of hyperons
in neutron star matter. The latter process has
been discussed in the literature
(Balberg \& Barnea \cite{bb98}, 
Schaab et al.\ \cite{sbs98}) and will be considered
briefly in Sect.\ 2.2.

Furthermore, in Eq.\ (\ref{Hw})
\begin{equation}
    l^\mu = \overline{\psi}_\nu \gamma^\mu (1 + \gamma^5) \psi_\nu
\label{l}
\end{equation}
is the neutrino 4-current ($\mu$=0,1,2,3),
$\gamma^\mu$ is a Dirac matrix,
$\psi_\nu$ is a neutrino bispinor amplitude,
\begin{equation}
     J_0 = \hat{\Psi}^+ \hat{\Psi},~~~
     {\vec J} = \hat{\Psi}^+ {\vec \sigma} \hat{\Psi}
\label{J}
\end{equation}
is the nucleon 4-current and ${\vec \sigma}$ is the Pauli vector matrix.

The nucleon current contains $\hat{\Psi}$, the secondary-quantized
nonrelativistic spinor wave
function of quasi-nucleons in superfluid matter,
and $\hat{\Psi}^+$, is its Hermitian conjugate.
$\hat{\Psi}$ is determined by the
Bogoliubov transformation. The transformation for
singlet-state pairing is well known (e.g., Lifshitz \&
Pitaevskii \cite{lp80}). The generalized Bogoliubov transformation
for a triplet-state $^3$P$_2$ pairing was studied in detail, for instance,
by Tamagaki (\cite{t70}). In the both cases $\hat{\Psi}$
can be written as
\begin{eqnarray}
  \hat{\Psi} & = & \sum_{{\vec p}\sigma \eta} \, \chi_\sigma \left[
               {\rm e}^{-iEt+i{\vec p}{\vec r}} \,
               U_{\sigma \eta}({\vec p}) \, \hat{\alpha}_{{\vec p}\eta} \right.
\nonumber \\
             &  + & \left.
               {\rm e}^{iEt-i{\vec p}{\vec r}} \,
               V_{\sigma \eta}(-{\vec p}) \, \hat{\alpha}_{{\vec p}\eta}^+
               \right],
\label{Psi}
\end{eqnarray}
where {\vec p} is a quasiparticle momentum,
\begin{equation}
   E=\sqrt{\epsilon^2 + \Delta_{\vec p}^2 }
\label{E}
\end{equation}
is its energy with respect to the Fermi level
(valid near the Fermi surface, at
$|p - p_{\rm F}| \ll p_{\rm F}$),
$\epsilon = v_{\rm F}(p-p_{\rm F})$,
$\sigma$ and $\eta=\pm 1$ enumerate spin states,
$\Delta_{\vec p}$ is a superfluid gap at the Fermi surface
($\Delta_{\vec p} \ll p_{\rm F} \, v_{\rm F}$),
$v_{\rm F}$ is the particle Fermi velocity;
$\chi_\sigma$ is a basic spinor
($\chi^+_\sigma \chi_{\sigma'} = \delta_{\sigma \sigma'} $),
$\hat{\alpha}_{{\vec p}\eta}$ and $\hat{\alpha}^+_{{\vec p}\eta}$ are,
respectively, the annihilation and creation operators.
$U_{\sigma\eta}({\vec p})$ and $V_{\sigma\eta}({\vec p})$
are matrix elements of the operators
$U({\vec p})$ and $V({\vec p})$ which realize the Bogoliubov transformation
from particle to quasiparticle states.
In the cases of singlet and triplet pairing,
one has
\begin{equation}
    U_{\sigma \eta}({\vec p}) = u_{\vec p} \, \delta_{\sigma \eta},~~~
    \sum_{\sigma \eta} |V_{\sigma \eta}({\vec p})|^2 =2 \,v_{\vec p}^2,
\label{UV}
\end{equation}
where
\begin{equation}
   u_{\vec p}  =  \sqrt{ {1 \over 2}
           \left( 1 + {\epsilon \over E} \right)},~~~
   v_{\vec p}  =  \sqrt{ {1 \over 2}
           \left( 1 - {\epsilon \over E} \right)}.
\label{uv}
\end{equation}
For a singlet-state pairing, the gap $\Delta_{\vec p}$
is actually independent of {\vec p}, so that $u_{\vec p}$
and $v_{\vec p}$ depend only on $p = |{\vec p}|$. For a
triplet-state pairing, $\Delta_{\vec p}$, $u_{\vec p}$ and
$v_{\vec p}$ depend on orientation of {\vec p}.
Note general symmetry properties (e.g., Tamagaki \cite{t70})
\begin{equation}
    U_{\sigma \eta}(-{\vec p}) = U_{\sigma \eta}({\vec p}),~~~
    V_{\sigma \eta}(-{\vec p}) = - V_{\eta \sigma}({\vec p}).
\label{symmetry}
\end{equation}
Let $q_\nu = (\omega_\nu,{\vec q}_\nu)$ and
$q'_\nu = (\omega'_\nu,{\vec q'}_\nu)$ be 4-momenta of
newly born neutrino and anti-neutrino, while
$p=(E,{\vec p})$ and $p'=(E',{\vec p}')$ be 4-momenta of annihilating
quasinucleons.
Using the Fermi Golden rule one can easily
obtain the neutrino emissivity in the form
\begin{eqnarray}
   Q & = & \left( {G_{\rm F} \over 2 \sqrt 2} \right)^2 \,
       {1 \over 2} \, {\cal N}_\nu \,
        \int {{\rm d} {\vec p} \over (2 \pi)^3 } \;
       {{\rm d} {\vec p}' \over (2 \pi)^3 } \;
       f(E)f(E')
\nonumber \\
     & \times &
     \int {{\rm d} {\vec q}_\nu \over 2 \omega_\nu (2 \pi)^3 } \;
     {{\rm d} {\vec q}'_\nu \over 2 \omega'_\nu (2 \pi)^3 } \;
     \left[c_V^2 I_{00} |l_0|^2 + c_A^2 \, I_{ik} l_i l_k^\ast \right]
\nonumber \\
     & \times &
     (2 \pi)^4 \, (\omega_\nu + \omega'_\nu) \,
     \delta^{(4)}(p + p' - q_\nu - q'_\nu) \,,
\label{Qgen}
\end{eqnarray}
where ${\cal N}_\nu$=3 is the number of neutrino flavors,
an overall factor $1/2$ is introduced to avoid
double counting of the same $\tilde{N}\tilde{N}$ collisions,
integration is meant to be carried out over
the domain $(q_\nu + q'_\nu)^2 >0$ in which the
process is kinematically allowed,
$f(E)=1/[\exp (E/T)+1]$ is the Fermi-Dirac distribution,
$T$ is the temperature,
$i,k$=1,2,3, and
\begin{eqnarray}
   I_{00} & = & \sum_{\eta\eta'} \,
   |\langle B | \hat{\Psi}^+ \hat{\Psi} | A \rangle |^2,
\nonumber \\
   I_{ik} & = & \sum_{\eta\eta'} \,
   \langle B | \hat{\Psi}^+ \sigma_i \hat{\Psi} | A \rangle \,
   \langle B | \hat{\Psi}^+ \sigma_k \hat{\Psi} | A \rangle^\ast.
\label{I}
\end{eqnarray}
In this case
$| A \rangle$ denotes an initial state of the quasinucleon system
(the individual states $({\vec p},\eta)$ and $({\vec p}',\eta')$ are occupied)
and $| B \rangle$ is a final state of the system
(the states $({\vec p},\eta)$ and $({\vec p}',\eta')$ are empty).
In Eq.\ (\ref{Qgen}) we neglected the interference terms
proportional to $c_V \, c_A$ since they vanish after
subsequent integration over {\vec p} and ${\vec p}'$.

Bilinear combinations of the neutrino current components (\ref{l})
are calculated in the standard manner, and integration over
d{\vec q}$_\nu$ and d{\vec q}$'_\nu$ is taken with the aid of the
Lenard integral. The result is:
\begin{eqnarray}
  &&   \int {{\rm d} {\vec q}_\nu \over 2 \omega_\nu} \;
     {{\rm d} {\vec q}'_\nu \over 2 \omega'_\nu} \;
     l^\alpha l^{\beta \ast} \,
     \delta^{(4)}(p + p' - q_\nu - q'_\nu)
\nonumber \\
   &&  =
     { 4 \pi \over 3} \, \left[q^\alpha q^\beta -
     (\omega^2 - {\vec q}^2 )\, g^{\alpha \beta} \right],
\label{ll}
\end{eqnarray}
where $q=(\omega,{\vec q})$ is 4-momentum of a
neutrino pair ($\omega=\omega_\nu + \omega'_\nu=E+E'$,
${\vec q}={\vec q}_\nu + {\vec q}'_\nu= {\vec p}+{\vec p}'$) and
$g^{\alpha \beta}$ is the metric tensor.
Inserting (\ref{ll}) into (\ref{Qgen}) we obtain
\begin{eqnarray}
   Q & = & \left( {G_{\rm F} \over 2 \sqrt 2} \right)^2 \,
       {2 \pi \over 3} \, { {\cal N}_\nu \over (2 \pi)^8}
       \int {\rm d} {\vec p} \;
       {\rm d} {\vec p}' \;
       f(E)f(E') \, \omega
\nonumber \\
  & \times &  \left\{ c_V^2  {\vec q}^2 I_{00}+
     c_A^2 \, \left[ q_i q_k I_{ik} +
     (\omega^2 - {\vec q}^2) \, I \right] \right\},
\label{Qgen1}
\end{eqnarray}
where $I=I_{xx}+I_{yy}+I_{zz}$.

Now the problem reduces to 6-fold integration over
quasinucleon momenta {\vec p} and ${\vec p}'$ within
the kinematically allowed domain $\omega^2 > {\vec q}^2$.
Since the nucleon Fermi liquid is assumed to be strongly
degenerate, only narrow regions of momentum space
near the nucleon Fermi-surface contribute into the reaction.
Thus we can set $p=p_{\rm F}$ and $p'=p_{\rm F}$
in all smooth functions under the integral.
One can prove that
the presence of superfluidity (of energy gaps)
makes the process kinematically
allowed in a small region of momentum space where
{\vec p} is almost antiparallel to ${\vec p}'$. This
allows us to set ${\vec p}'=-{\vec p}$ in all smooth
functions in the integrand.

For further integration in Eq.\ (\ref{Qgen1}) we write
${\rm d}{\vec p} = p_{\rm F}^2 \, {\rm d}k \, {\rm d}\Omega$ and
${\rm d}{\vec p}' = p_{\rm F}^2 \, {\rm d}k' \, {\rm d}\Omega'$,
where d$\Omega$ and d$\Omega'$ are solid angle elements,
$k=p - p_{\rm F}$ and $k'=p' - p_{\rm F}$. Let us integrate
over ${\rm d}\Omega'$ first. For this purpose, we can
fix {\vec p} and introduce a local reference frame
$XYZ$ with the $Z$-axis antiparallel to {\vec p}. Let $\Theta$
and $\Phi$ be, respectively, azimuthal and polar angles
of ${\vec p}'$ with respect to $XYZ$. Since the
space allowed for ${\vec p}'$ is small,
we have $\Theta \ll 1$, $q_X = q_\perp \cos \Phi$,
$q_Y = q_\perp \sin \Phi$, $q_Z = k'-k \equiv q_\parallel $, where
$q_\perp = p_{\rm F} \sin \Theta \approx p_{\rm F} \, \Theta$. In this case
${\rm d}{\vec p}' = {\rm d}k' \; q_\perp \, {\rm d}q_\perp \;
{\rm d}\Phi$, ${\vec q}^2 = |{\vec p} + {\vec p}' |^2
=q_\parallel^2+q_\perp^2$,
and $\omega^2 - {\vec q}^2 = q_{\perp 0}^2 - q_\perp^2 \geq 0$,
where $q_{\perp0}^2 = \omega^2 - q_\parallel^2$. The quantities
$I_{00}$ and $I_{ik}$ are smooth functions of {\vec p} and
${\vec p}'$. In these functions, we set ${\vec p}'= - {\vec p}$
which makes them independent of {\vec q}. Therefore,
$q_X$ and $q_Y$ are the only variables which depend on $\Phi$.
The integration over $\Phi$ in Eq.\ (\ref{Qgen1})
contains the term,
\begin{eqnarray}
   \int_0^{2\pi} \,I_{ik} \, q_i q_k \,{\rm d\Phi} =
   \pi \left[ 2 q_\parallel^2 I_{ZZ} + q_\perp^2 \, (I_{XX}+I_{YY}) \right] &&
\nonumber \\
  = \,  \pi \left[ 2 q_\parallel^2 n_i n_k \, I_{ik} + q_\perp^2
    (\delta_{ik} - n_i n_k) \, I_{ik} \right], &&
\label{intPhi}
\end{eqnarray}
where ${\vec n} = {\vec p}/p$. Integration is performed
in the local coordinate frame $XYZ$, but the result
is transformed to the basic coordinate frame using
tensor character of $I_{ik}$.
Subsequent integration over $q_\perp$ from 0 to $q_{\perp 0}$
is easy and yields
\begin{eqnarray}
   Q & = & \left( {G_{\rm F} \over 2 \sqrt 2} \right)^2 \,
       {\pi^2  p_{\rm F}^2 \over 6} \, { {\cal N}_\nu \over (2 \pi)^8}
       \int {\rm d}\Omega \; \int \int
       {\rm d}k \, {\rm d}k' \;
       f(E)f(E') 
\nonumber \\
      & \times & \omega \,
       (\omega^2 - q_\parallel^2)
     \left\{ 2 c_V^2  (\omega^2+q_\parallel^2) I_{00}
     \right.
\nonumber \\
     & + & \left. g_V^2 \, \left[  (5 q_\parallel^2 - \omega^2)
     \, I_{ik} n_i n_k  +
      3 \, (\omega^2 - q_\parallel^2) \, I \right] \right\}.
\label{Qgen2}
\end{eqnarray}
Now we introduce dimensionless variables
\begin{equation}
  x = { v_{\rm F} k \over T},~~~
  x' = { v_{\rm F} k' \over T},~~~
  z = {E \over T},~~~z' = {E' \over T},
\label{var}
\end{equation}
which give
\begin{eqnarray}
   Q & = & \left( {G_{\rm F} \over 2 \sqrt 2} \right)^2 \,
       {\pi^2  p_{\rm F}^2 \over 6 v_{\rm F}^2 } \,
       { {\cal N}_\nu \, T^7 \over (2 \pi)^8}
       \int {\rm d}\Omega \; \int \int
       {\rm d}x \, {\rm d}x' \;
\nonumber \\
      & \times &  { (z + z') \over ({\rm e}^z + 1) ({\rm e}^{z'} + 1 )}
       \left[ (z+z')^2 - {(x-x')^2 \over v_{\rm F}^2} \right]
\nonumber \\
      & \times &
      \left\{ 2 c_V^2  \left[(z+z')^2+
      {(x-x')^2 \over v_{\rm F}^2} \right] \, I_{00} \right.
\nonumber \\
     & + & c_A^2 \, \left[
     3 \, \left( (z+z')^2 -
     { (x-x')^2 \over v_{\rm F}^2 }  \right) \, I \, \right.
\nonumber \\
      & - & \left. \left.
     \left( (z+z')^2 - {5\, (x-x')^2 \over v_{\rm F}^2 } \right) \,
     I_{ik} n_i n_k  \right] \right\},
\label{Qgen3}
\end{eqnarray}
and the integration over $x$ and $x'$ is restricted by the
domain where $(z+z')>|x-x'|/v_{\rm F}$. The outer integration
is over orientations of nucleon momentum {\vec p}.
Performing the inner
integration over $x$ and $x'$ we can assume that this orientation
is fixed (and the vector {\vec n} is constant). Then
the superfluid gap $\Delta_{\vec p}$ is fixed as well.
Introducing $y=\Delta_{\vec p}/T$ we have $z=\sqrt{x^2 + y^2}$ and
$z'=\sqrt{x'^2 + y^2}$. The integration domain can be rewritten
as $(x-x')^2 - v_{\rm F}^2 (x+x')^2 \leq
4 v_{\rm F}^2 y^2 /(1-v_{\rm F}^2)$.
In the nonrelativistic limit, we are interested in,
$v_{\rm F} \ll 1$, and the domain transforms to the narrow
strip in the $xx'$ plane near the $x=x'$ line.
It is sufficient to set $x'=x$ and $z'=z$ in smooth functions
and integrate over
$\delta x = x'-x$ in the narrow range $|\delta x| \leq 2zv_{\rm F}$.
In this way we come to a simple equation
\begin{eqnarray}
  &&   \int_{-2zv_{\rm F}}^{2zv_{\rm F}} {\rm d}\delta x
       \left[(z+z')^2 - {(x-x')^2 \over v_{\rm F}^2} \right]
\nonumber \\
  & \times &   \left\{ 2 c_V^2  \left[(z+z')^2+
     {(x-x')^2 \over v_{\rm F}^2} \right] \,  I_{00} \right.
\nonumber \\
     & + & c_A^2 \, \left[
     3 \, \left( (z+z')^2 -
     { (x-x')^2 \over v_{\rm F}^2 }  \right) \, I \, \right.
\nonumber \\
      & - & \left. \left.
     \left( (z+z')^2 - {5\, (x-x')^2 \over v_{\rm F}^2 } \right) \,
     I_{ik} n_i n_k  \right] \right\}
\nonumber \\
   & = & {2^9 \over 5} z^5 v_{\rm F} \, (c_V^2 I_{00} + c_A^2 I),
\label{intdx}
\end{eqnarray}
and the most sophisticated term containing $I_{ik}n_i n_k$
vanishes.

\subsection{Practical formulae}

Inserting (\ref{intdx}) into (\ref{Qgen3}) and returning to
the standard physical units we have
\begin{eqnarray}
 Q & = & {4 G_{\rm F}^2
           m_N^\ast p_{\rm F} \over 15 \pi^5 \hbar^{10}
           c^6} \, (k_{\rm B} T)^7 \, {\cal N}_\nu R =
\nonumber \\
     & = & 1.170 \times 10^{21} \,  {m_N^\ast \over m_N } \;
       { p_{\rm F} \over m_N c   }  \, T_9^7 \, {\cal N}_\nu \, R~~~
      {{\rm erg} \over {\rm cm^3 \; s  }},
\label{Q}
\end{eqnarray}
where $m_N^\ast=p_{\rm F}/v_{\rm F}$ is an effective quasinucleon mass,
$m_N$ is bare nucleon mass, $T_9= T/(10^9 \,{\rm K})$,
$k_{\rm B}$ is the Boltzmann constant,
and
\begin{eqnarray}
    R & = & {1 \over 8 \pi} \, \int {\rm d}\Omega \,
     \int_0^\infty \, {z^6 \, {\rm d}x \over ({\rm e}^z +1)^2} \,
    (c_V^2 \, I_{00} + c_A^2 \, I)
\label{R}
\end{eqnarray}
is a function to be determined.
The integrand contains the functions $I_{00}$ and
$I=I_{xx}+I_{yy}+I_{zz}$ defined by Eqs.\ (\ref{I}) with
${\vec p}'=-{\vec p}$ (see above). From Eqs.\ (\ref{UV})--(\ref{symmetry})
for the singlet- and triplet-state pairings
we obtain the same expression
\begin{equation}
    I_{00}=8 u_{\vec p}^2 v_{\vec p}^2 = 2 { y^2 \over z^2}.
\label{I00}
\end{equation}
In the case of the singlet-state pairing the Bogoliubov
operator $V({\vec p})$ possesses the properties
$V_{\alpha \beta}(-{\vec p})= V_{\alpha \beta}({\vec p})$ and
$V_{\alpha \beta}({\vec p})=-V_{\beta \alpha}({\vec p})$,
and has the form (e.g, Tamagaki \cite{t70})
\begin{equation}
    V_{\alpha \beta}({\vec p}) = \left(
     \begin{array}{cc}
        0        & v_{\vec p}   \\
     -v_{\vec p}  &   0
     \end{array}
     \right).
\end{equation}
Then from Eqs.\ (\ref{I}) we have $I=0$, i.e.,
the axial-vector contribution vanishes for the singlet-state
pairing in accordance with the result by Flowers et al.\ (\cite{frs76})
(to be exact, the main term in the
nonrelativistic expansion of $Q$ over $(v_{\rm F}/c)^2$
vanishes).
In this case the gap is isotropic
and integration over d$\Omega$ is trivial:
\begin{equation}
    R = c_V^2 F_s,~~~F_s=
    y^2 \int_0^\infty \, {z^4 \, {\rm d}x \over ({\rm e}^z +1)^2}.
\label{Rs}
\end{equation}

For the triplet-state pairing, according to
Tamagaki (\cite{t70}), 
$V_{\alpha \beta}(-{\vec p})= - V_{\beta \alpha}({\vec p})$,
$V_{\alpha \beta}({\vec p})=V_{\beta \alpha}({\vec p})$, and
$V_{\alpha \beta}({\vec p})= v_{\vec p} \Gamma_{\alpha \beta}({\vec p})$,
where $\Gamma_{\alpha \beta}({\vec p})$
is a unitary (2$\times$2) matrix. Using these relationships and
Eq.\ (\ref{UV}), for the triplet case
from (\ref{I}) we obtain
\begin{equation}
    I = 8 u_{\vec p}^2 \
    \sum_{\alpha \beta} |V_{\alpha \beta}({\vec p})|^2
     = 16 u_{\vec p}^2 v_{\vec p}^2 = 4 \, {y^2 \over z^2},
\label{It}
\end{equation}
which yields
\begin{eqnarray}
    R & = & (c_V^2 + 2 \, c_A^2)\,F_t,
\nonumber \\
     F_t & = & {1 \over 4 \pi} \, \int {\rm d}\Omega \, y^2
     \int_0^\infty \, {z^4 \, {\rm d}x \over ({\rm e}^z +1)^2}.
\label{Rt}
\end{eqnarray}
Contrary to the singlet-state paring, the axial-vector
contribution does not vanish. 

The results (\ref{Rs}) and (\ref{Rt}) for
$^1$S$_0$ and $^3$P$_2$ superfluids can be written
in a unified manner:
\begin{equation}
    R = a \, F,
\label{RR}
\end{equation}
where $F$ stands for $F_s$ or $F_t$, while $a=c_V^2$ or
$c_V^2+ 2 \, c_A^2$ is a dimensionless reaction constant
that depends on the particle species and superfluid type.
In Table 1 we list the values of $a$ for singlet-state and
triplet-state superfluids of neutrons and protons
(calculated using the values of
$c_V$ and $c_A$ cited in Sect.\ 2.1).
(A) denotes $^1$S$_0$ pairing, while (B) and (C)
are two types of $^3$P$_2$ pairing with total
projection of the Cooper--pair momentum onto the $z$-axis equal
to $m_J=0$ and 2, respectively.
One can hardly
expect triplet-state pairing of protons in a neutron star core
but we present the corresponding value for completeness of discussion.
We give also the values of $a$ for singlet-state pairing
of hyperons. Hyperon superfluidity has been discussed recently by
Balberg \& Barnea (\cite{bb98}) and incorporated into
calculations of the neutron star cooling by Schaab et al.\ (\cite{sbs98}).
The value of $a$ for hyperons is determined by the vector constant
$c_V$ of weak neutral currents in Eq.\ (\ref{Hw})
as a sum of contributions of corresponding quarks
(e.g., Okun' \cite{o90}). 
One can see that the efficiency of the neutrino emission
due to singlet-state pairing of various particles is drastically
different. The emission is quite open for
$n$, $\Sigma^\pm$, $\Xi^0$ but strongly (by two orders of
magnitude) reduced for $p$ and $\Xi^-$,
and vanishes for $\Lambda$ and $\Sigma^0$ hyperons.
Notice that the values of $c_V$, $c_A$ and $a$ can be renormalized
by manybody effects in dense matter which we ignore, for simplicity.
Notice also that in the cases
of singlet--state pairing of protons and $\Xi^-$
the first non-vanishing relativistic corrections
to the emissivity $Q$ due to axial--vector neutral currents
($ \sim v_{\rm F}^2$) could be larger than the small ($ \sim c_V^2 $)
zero--order contribution of the vector currents.
Since we neglect relativistic corrections,
our expressions for $Q$ in these cases may be somewhat inaccurate
(give reliable lower limits of $Q$).

\begin{figure}[htb]
\begin{center}
\leavevmode
\epsfysize=8.5cm \epsfbox{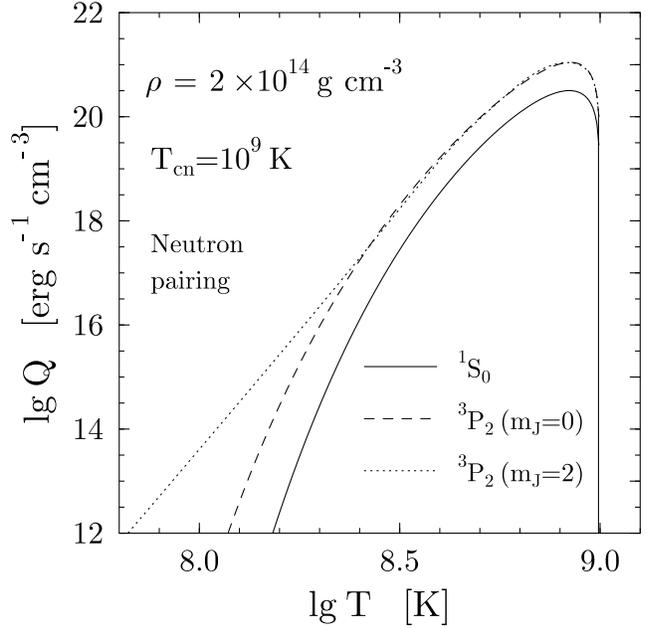}
\end{center}
\caption[]{
Temperature dependence of the neutrino emissivity
due to Cooper pairing of neutrons
at $\rho = 2 \times 10^{14}$ g cm$^{-3}$
and $T_{cn} = 10^9$ K for superfluidity (A) (solid line),
(B) (dashed line) and (C) (dots).
}
\label{fig-Q}
\end{figure}

\begin{table}[ht]
\caption{Reaction constant $a$ in Eq.\ (\protect{\ref{RR}})   }
\label{tab1}
\begin{tabular}{lll}
\hline
\noalign{\smallskip}
\hline
\noalign{\smallskip}
particles   &    pairing & $a$ \\
\noalign{\smallskip}
\hline
\noalign{\smallskip}
 $n$, $\Xi^0$                         & $^1$S$_0$ (A)    &  1    \\
 $n$                                  & $^3$P$_2$ (B,C)  &
 $1+2g_A^2=4.17$ \\
 $p$, $\Xi^-$                         & $^1$S$_0$ (A)    &
 $(1-4 \sin^2 \Theta_{\rm W})^2=0.0064$ \\
 $p$                                  & $^3$P$_2$ (B,C)  &
 $(1-4 \sin^2 \Theta_{\rm W})^2+2 g_A^2=3.18$ \\
 $\Sigma^\pm$                              & $^1$S$_0$ (A)    &
 $(2 - 4 \sin^2 \Theta_{\rm W})^2=1.17$ \\
 $\Lambda$, $\Sigma^0$                 & $^1$S$_0$ (A)  &   0 \\
\noalign{\smallskip}
\hline
\noalign{\smallskip}
\hline
\end{tabular}
\end{table}

Equations (\ref{Q}) and (\ref{Rs}) for singlet-state pairing
of neutrons with
two neutrino flavors (${\cal N}_\nu=2$) were obtained by
Flowers et al.\ (\cite{frs76}).
Similar equations were derived by
Voskresensky \& Senatorov (\cite{vs86, vs87}). Note that the final
expressions for $Q$ obtained by the latter authors
contain a misprint: there is $\pi^2$ instead of $\pi^5$
in the denominator although
the numerical formula includes the correct factor $\pi^5$.
In addition, the expressions by
Voskresensky \& Senatorov (\cite{vs86, vs87}) are written for one neutrino
flavor and erroneously contain the
axial-vector contribution which is actually negligible.

Now we obtain practical expressions for the function
$F$ in Eq.\ (\ref{RR}). Following Levenfish \&
Yakovlev (\cite{ly94a}, b) we consider three types of BCS superfluid:
(A), (B) and (C) as described above.
In case (A) the superfluid gap is
isotropic, $\Delta_A = \Delta_0^{(A)}(T)$. In cases (B) and (C) the gap
is anisotropic and depends on angle $\theta$ between
quasinucleon momentum {\vec p} and the $z$-axis:
$\Delta_B=\Delta_0^{(B)}(T)\, \sqrt{1 + 3 \cos^2 \theta}$,
$\Delta_C=\Delta_0^{(C)}(T)\,\sin \theta $, respectively, where
$\Delta_0(T)$ is a temperature-dependent amplitude.
Therefore, at given $T$ the gap $\Delta_B$ has minimum equal
to $\Delta_0^{(B)}(T)$
for quasinucleons at the equator of the Fermi-sphere,
whereas the gap $\Delta_C$ has maximum $\Delta_0^{(C)}(T)$
at the equator and nodes at the poles of the Fermi-sphere.
In the cooling theories of neutron stars one
commonly considers the nodeless pairing (B)
of neutrons. However,
thermodynamics of nucleon superfluid
is very model-dependent and one cannot
exclude that the C-type superfluid appears in the
neutron star cores instead of the B-type, at least
at some temperatures and densities.

Let us introduce the notations
\begin{equation}
  v = {\Delta_0(T) \over  k_{\rm B}T},~~~~~
  \tau = { T \over T_c},
\label{vtau}
\end{equation}
where $T_c$ is the critical temperature of nucleon superfluidity.
Using the standard equations of the BCS theory,
Levenfish \& Yakovlev (\cite{ly94a}, b) obtained analytic fits
which relate $\Delta_0(T)$ to $\tau$ at any $\tau < 1$
for all three superfluid types:
\begin{eqnarray}
    v_A &=& \sqrt{1-\tau}
            \left( 1.456 - \frac{0.157}{\sqrt{\tau}} + \frac{1.764}{\tau}
            \right),
\nonumber \\
    v_B &=& \sqrt{1-\tau} \left (0.7893 + \frac{1.188}{\tau} \right),
\nonumber \\
    v_B &=& \frac{\sqrt{1-\tau^4}}{\tau}
              \left(2.030 - 0.4903  \, \tau^4 + 0.1727 \, \tau^8 \right).
\label{v-tau}
\end{eqnarray}
%

\begin{figure}[htb]
\begin{center}
\leavevmode
\epsfysize=8.5cm 
\epsfbox{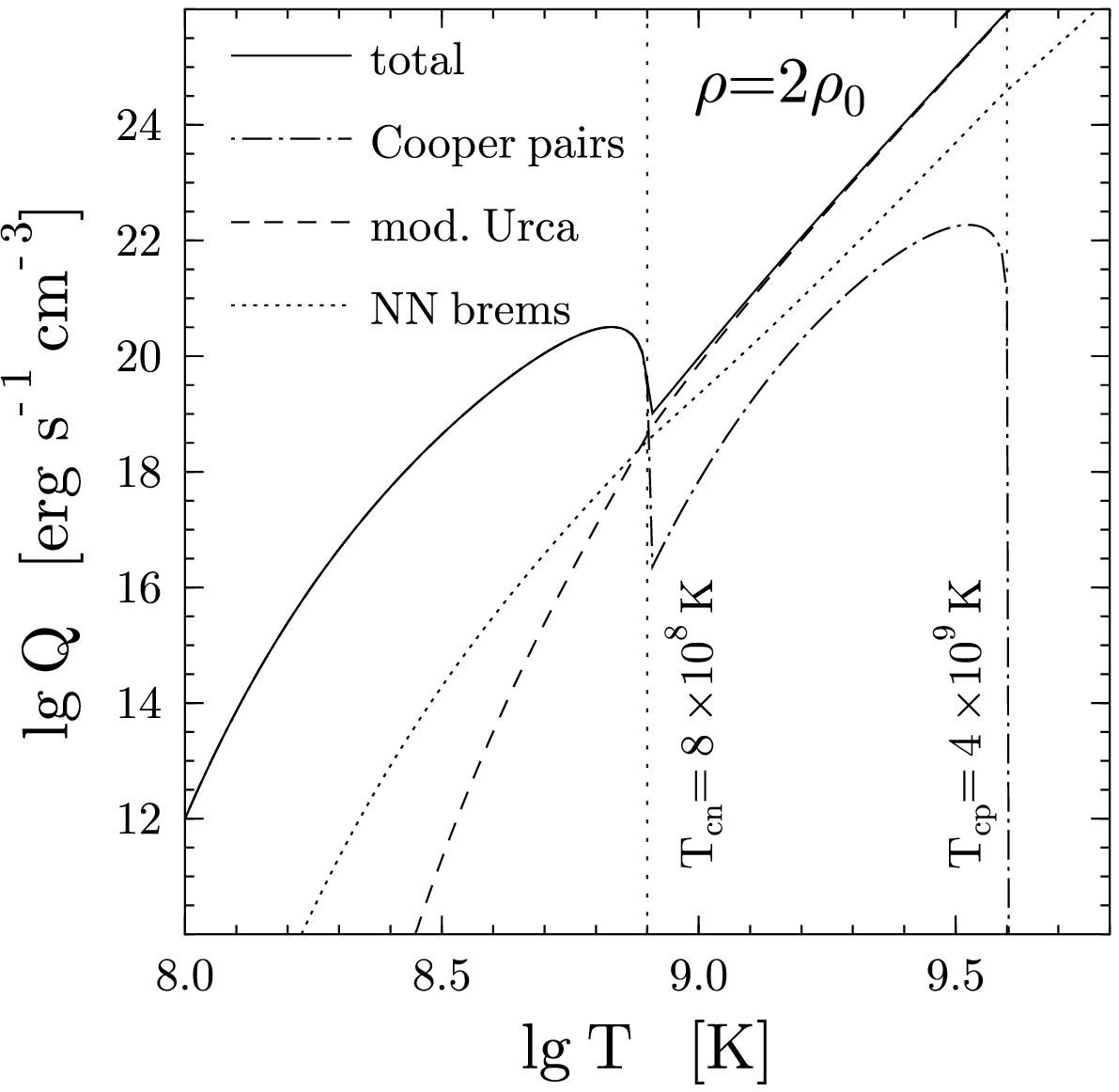}
\end{center}
\caption[]{
Temperature dependence of the neutrino emissivities
in different reactions at $\rho = 2 \rho_0$
for neutron superfluid (B) with $T_{cn} = 8 \times 10^8$ K,
and proton superfluid (A) with $T_{cp} = 4 \times 10^9$~K
}
\label{fig2rho}
\end{figure}
%

\begin{figure}[thb]
\begin{center}
\leavevmode
\epsfysize=8.5cm
\epsfbox{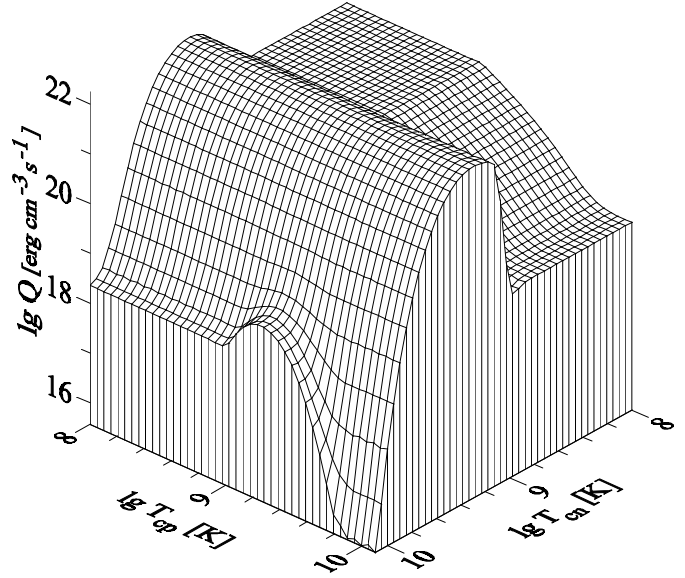}
\end{center}
\caption[]{
Total neutrino emissivity versus $T_{cn}$ and $T_{cp}$
in a neutron star core at $\rho = 2 \rho_0$ and $T=10^9$ K.
}
\label{fig2rho2}
\end{figure}

One can easily see that the function $F$
in Eq.\ (\ref{RR}) depends actually on the only parameter $v$ and
on the superfluid type.
An analysis of this function
from Eqs.\ (\ref{Rs}) and (\ref{Rt})
is quite similar to that carried out by Levenfish \&
Yakovlev (\cite{ly94b}) in their study of the effect of superfluidity
on the heat capacity of nucleons. Therefore we will omit
technical details and present the final results.

Just after the superfluidity onset when
the dimensionless gap parameter
$v \ll 1$ and $\tau = T/T_c \to 1$, we obtain
\begin{eqnarray}
   F_A(v)&  = & 0.602 \, v^2 = 5.65 \, (1-\tau),
\nonumber \\
   F_B(v)&  = & 1.204 \, v^2 = 4.71 \, (1-\tau),
\nonumber \\
   F_C(v)&  = & 0.4013 \, v^2 = 4.71 \, (1-\tau).
\label{low-v}
\end{eqnarray}
At temperatures $T$ much below $T_c$ one has $v \gg 1$ and
\begin{eqnarray}
   F_A(v) & = & {\sqrt{\pi} \over 2} \, v^{13/2} \,
     {\rm e}^{-2v} = {35.5 \over \tau^{13/2}} \, {\rm e}^{-2v},
\nonumber \\
   F_B(v) & = & {\pi \over 4 \sqrt{3}} \, v^6 \,
     {\rm e}^{-2v} = {1.27 \over \tau^6} \, {\rm e}^{-2v},
\nonumber \\
   F_C(v) & = & {50.03 \over v^2} = 12.1 \, \tau^2.
\label{high-v}
\end{eqnarray}
Therefore the neutrino emission due to the nucleon pairing
differs significantly from the majority of other
neutrino reactions. The process has a threshold
(becomes allowed at $T \leq T_c$), and the neutrino
emissivity $Q$ is a nonmonotonic function of temperature.
It grows rapidly with decreasing $T$ just below $T_c$
which does not happen in other reactions.
With further decrease of $T$ the emissivity
$Q$ reaches maximum and then decreases.
According to Eqs.\ (\ref{high-v})
the decrease of $Q$ is exponential for the nodeless
superfluids of (A) or (B), and it is
power-law ($Q \propto T^9$) for the superfluid (C).
The power-law behaviour of $Q$ in case (C)
occurs due to the presence of nodes in the
superfluid gap. At $T \ll T_c$
superfluids (A), (B), and (C)
suppress the Cooper-pair neutrino emission
in the same manner in which they suppress
the heat capacity and the direct Urca process
(Levenfish \& Yakovlev \cite{ly94a}, b).

Finally, we have calculated $F(v)$
numerically in a wide range of
$v$ and fitted the results by simple expressions
which reproduce also the asymptotes
(\ref{low-v}) and (\ref{high-v}):
\begin{eqnarray}
  F_A(v) & = &  (0.602 \, v^2 + 0.5942\, v^4 +
     0.288 \, v^6)
\nonumber \\
       & \times &
     \left( 0.5547 + \sqrt{(0.4453)^2 + 0.01130 \,v^2} \right)^{1/2}
\nonumber \\
     & \times &
     \exp \left(- \sqrt{4 \, v^2 + (2.245)^2 } + 2.245 \right),
\nonumber \\
  F_B(v) & = & {1.204 \, v^2 + 3.733 \, v^4 +
     0.3191 \, v^6 \over 1 + 0.3511 \, v^2}
\nonumber \\
   & \times & \left( 0.7591 + \sqrt{ (0.2409)^2 + 0.3145 \, v^2 } \right)^2
\nonumber \\
    & \times &
    \exp \left( - \sqrt{ 4 \, v^2 + (0.4616)^2} + 0.4616 \right),
\nonumber \\
  F_C(v) & = &
    (0.4013 \, v^2 - 0.043 \, v^4 + 0.002172 \, v^6)
\nonumber \\
     & \times &
       ( 1 - 0.2018 \, v^2
       +   0.02601 \, v^4
\nonumber \\
     & - & 0.001477 \, v^6 + 0.0000434 \, v^8)^{-1}.
\label{fit}
\end{eqnarray}
The maximum fit error is
about 1\% at $v \approx 4$ for $F_A(v)$;
about 3.4\% at $v \approx 2$ for $F_B(v)$;
and about 3\% at $ v \approx 1$ for $F_C(v)$.

%
\section{Efficiency of Cooper-pair neutrinos}

Equations (\ref{Q}), (\ref{RR})
and (\ref{fit}) enable one to calculate
the neutrino emissivity $Q$ due to Cooper
pairing of nucleons or hyperons.
For illustration, we use a moderately stiff
equation of state in a neutron star core proposed
by Prakash et al.\ (\cite{pal88})
(the version with the compression modulus $K_0=180$ MeV,
and with the same simplified symmetry factor
$S_V$ that has been used by Page \& Applegate, \cite{pa92}).
According to this equation of state,
dense matter consists of neutrons with admixture of protons and
electrons (no hyperons).
We set the effective nucleon masses
$m_N^\ast = 0.7 \, m_N$, for simplicity.
Since the critical temperatures $T_{cn}$ and $T_{cp}$
of the neutron and proton superfluids are model dependent
(e.g., Tamagaki \cite{t70}, Amundsen \&  {\O}stgaard \cite{ao85},
Baldo et al.\ \cite{bcll92},
Takatsuka \& Tamagaki \cite{tt93}, \cite{tt97}),
we do not use any specific microscopic superfluid model
but treat $T_{cn}$ and $T_{cp}$ as free parameters.

\begin{figure}[htb]
\begin{center}
\leavevmode
\epsfysize=8.5cm
\epsfbox{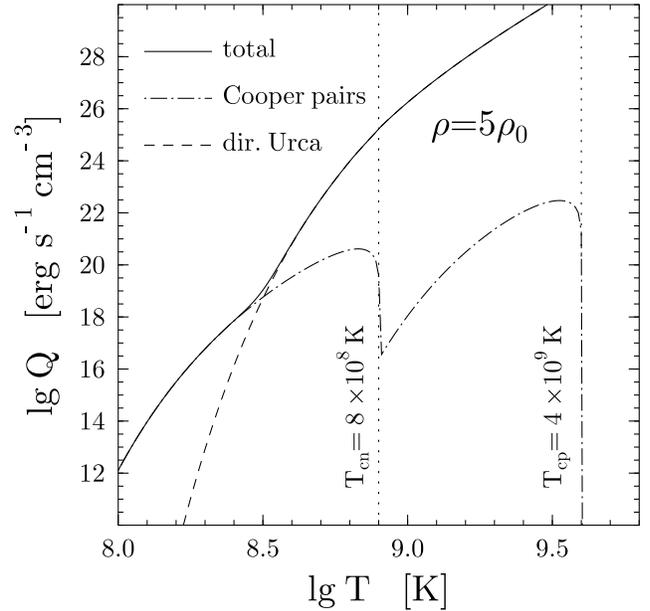}
\end{center}
\caption[]{
Temperature dependence of the neutrino emissivities
in main reactions
at $\rho = 5 \rho_0$
for neutron superfluid (B) with $T_{cn} = 8 \times 10^8$ K,
and proton superfluid (A) with $T_{cp} = 4 \times 10^9$~K
}
\label{fig5rho}
\end{figure}

Figure \ref{fig-Q} shows temperature dependence of the emissivity
$Q$ produced by Cooper pairing of neutrons
in the neutron star core at
$\rho = 2 \times 10^{14}$ g cm$^{-3}$. The critical temperature
is assumed to be $T_{cn} = 10^9$ K. Given $\rho$ is typical for
the transition from the singlet-state
to the triplet-state pairing (Sect.\ 1). Thus various superfluid types
are possible according to different microscopic theories.
We present the curves for all three
superfluid types (A), (B) and (C) discussed in Sect.\ 2.2.
A growth of the emissivity with decreasing $T$
below $T_c$ is very steep. The main neutrino emission
occurs in the temperature range
$0.4 \, T_c \la T \la T_c$. In this range,
the emissivity depends weakly on the superfluid type
and is rather high, of the order of or even higher
than in the modified Urca reaction (\ref{Murca}) in non--superfluid
matter. This indicates that Cooper-pair neutrinos
are important for neutron star cooling (Sect.\ 4).

%
\begin{figure*}[htb]
\begin{center}
\leavevmode
\epsfysize=8.5cm
\epsfbox{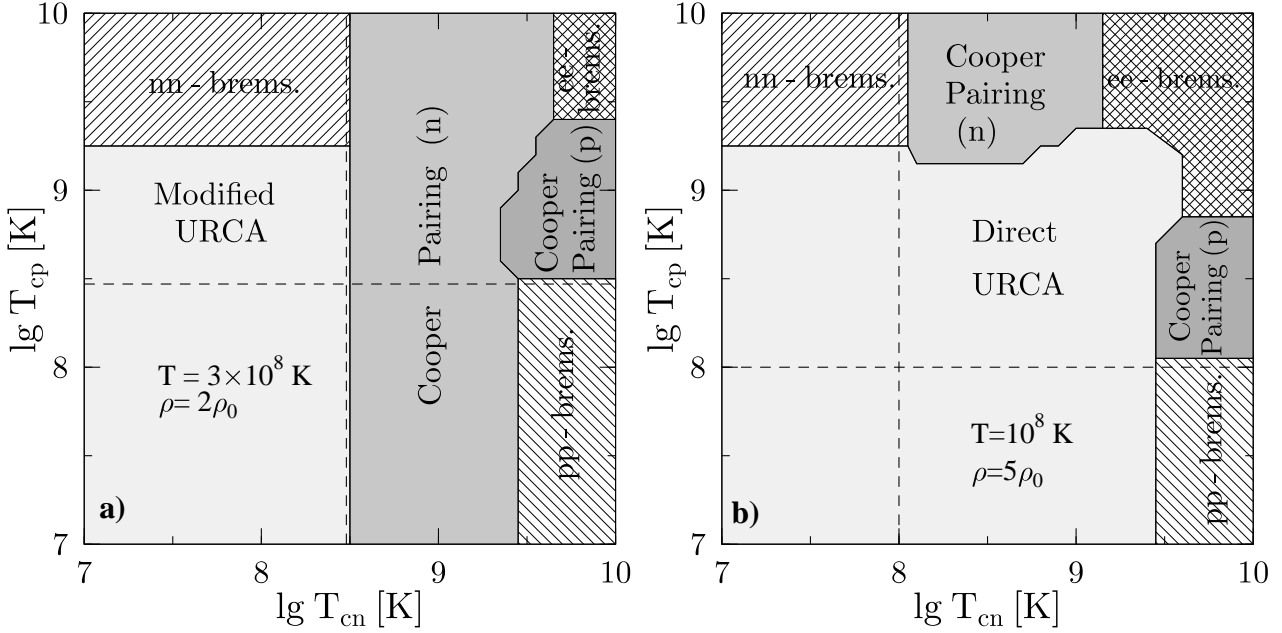}
\end{center}
\caption[]{Domains of $T_{cp}$ and $T_{cn}$, where different
           neutrino emission mechanisms dominate
           (different shaded regions).
           In addition to the mechanisms
           (\protect{\ref{Murca}})--(\protect{\ref{CPF}}) we include
           also the neutrino
           emission due to $ee$ bremsstrahlung
           (Kaminker et al.\ \cite{kyh97}).
           Figure (a) corresponds to the standard neutrino
           emission at $\rho = 2 \, \rho_0$ and
           $T= 3 \times 10^8$ K, while Figure (b)
           refers to the emission enhanced by the direct Urca
           process at
           $\rho= 5\, \rho_0$ and $T = 10^8$ K.
           Dashed lines correspond to $T_{cn}=T$ or $T_{cp}=T$.
           }
\label{qq_reca}
\end{figure*}

Figure \ref{fig2rho} shows temperature dependence of
various neutrino energy generation rates in a neutron
star core at $\rho = 2 \rho_0$. The neutron critical temperature is
assumed to be $T_{cn} = 8 \times 10^8$ K,
while the proton critical temperature is
$T_{cp} = 4 \times 10^9$ K. We show the contributions
from the modified Urca reaction (\ref{Murca})
(sum of the proton and neutron branches),
nucleon-nucleon bremsstrahlung (\ref{Nbrems})
(sum of $nn$, $np$ and $pp$ contributions),
Cooper pairing of nucleons (sum of $n$ and $p$ reactions),
and also the total neutrino emissivity (solid line).
The direct Urca process (\ref{Durca}) is forbidden at given $\rho$.
Here and in what follows the rates of the reactions
(\ref{Murca})--(\ref{Durca})
are taken as described in Levenfish \& Yakovlev (\cite{ly96}),
with proper account of suppression of the reactions by
neutron and/or proton superfluidity (Levenfish \&
Yakovlev \cite{ly94a}, Yakovlev \& Levenfish \cite{yl95}).
The neutron superfluid is assumed
to be of type (B) and the proton superfluid of type (A).
A large bump of the total emissivity at
$T \approx 10^{8.8}$ K is produced by the Cooper pairing of
neutrons. If neutrino emission due to this pairing were absent
the total neutrino emissivity at $T \la T_{cn}$ would
be 2 -- 4 orders of magnitude smaller owing to the strong suppression
by the nucleon superfluidity. The Cooper pairing
can easily turn suppression into enhancement.

Figure \ref{fig2rho2} shows a joint effect of the neutron and
proton superfluids onto the total neutrino emissivity $Q$
at $\rho = 2 \rho_0$ and
$T= 10^9$ K. Presentation of $Q$ as a function of
$T_{cn}$ and $T_{cp}$ allows us to display all the effects
of superfluidity onto the neutrino emission.
If $T>T_{cn}$ and $T>T_{cp}$
we have $Q$ in nonsuperfluid
matter (a plateau at small $T_{cn}$ and $T_{cp}$).
For other $T$, the neutron and/or proton superfluidity
affects the neutrino emission. It is seen that the neutron pairing at
$T \la T_{cn}$ may enhance the neutrino energy losses.
A similar effect of the proton pairing at $T \la T_{cp}$
is much weaker as explained above. In a strongly superfluid matter
(highest $T_{cn}$ and $T_{cp}$)
the neutrino emission is drastically suppressed
by the superfluidity.

Figure \ref{fig5rho} shows temperature dependence
of some partial and total neutrino emissivities
for the same $T_{cn}$ and $T_{cp}$ as in Fig.\ \ref{fig2rho}
but for higher density $\rho = 5 \rho_0$.
The equation of state we adopt opens the powerful direct Urca
process at
$\rho >  \rho_{cr} =  1.30 \times  10^{15}$ g cm$^{-3}=4.64 \, \rho_0$.
We take into account all major neutrino
generation reactions (\ref{Murca})--(\ref{CPF}).
For simplicity, we do not show all partial emissivities
since the total emissivity is mainly
determined by the
interplay between the direct Urca and Cooper pairing reactions.

The effect of Cooper-pair neutrinos is seen to be less important
than for the standard neutrino reactions in Fig.\ \ref{fig2rho}
but, nevertheless quite noticeable. It is especially
pronounced if $T \la T_{cn} \ll T_{cp}$. In this case,
the strong proton superfluid suppresses greatly
the direct Urca process, and the emission due to
Cooper pairing of neutrons can be significant.

Finally notice that while calculating neutron star
cooling one often assumes the
existence of one dominant neutrino generation mechanism,
for instance, the direct Urca process for the enhanced cooling
or the modified Urca process for the standard cooling.
This is certainly true for non--superfluid neutron--star cores,
but wrong in superfluid matter. In the latter case,
different mechanisms can dominate (Fig.\ \ref{qq_reca})
at various cooling stages depending on temperature,
$T_{cn}$, $T_{cp}$, and density.

In particular, the Cooper-pair neutrinos dominate
the standard neutrino energy losses at $T \! \la \! 10^9$~K
for a moderate neutron superfluidity ($0.12 \! \la T/T_{cn} \la \! 0.96$).
This parameter range is important for cooling theories.
At the early cooling stages, when
$T \ga 10^9$~K,
the Cooper-pair neutrinos can also be important
but in a narrower range around
$T/T_{cn}  \! \approx 0.4$ especially in the presence
of the proton superfluid.
As mentioned above,
Cooper-pair neutrinos can dominate also in the
regime of rapid neutrino emission
if the nucleons of one species are strongly superfluid but
the other nucleons are moderately or weakly superfluid
(see Fig.\ \ref{qq_reca}b).

%
\begin{figure*}[htb]                         
\begin{center}
\leavevmode
\epsfysize=8.5cm
\epsfbox{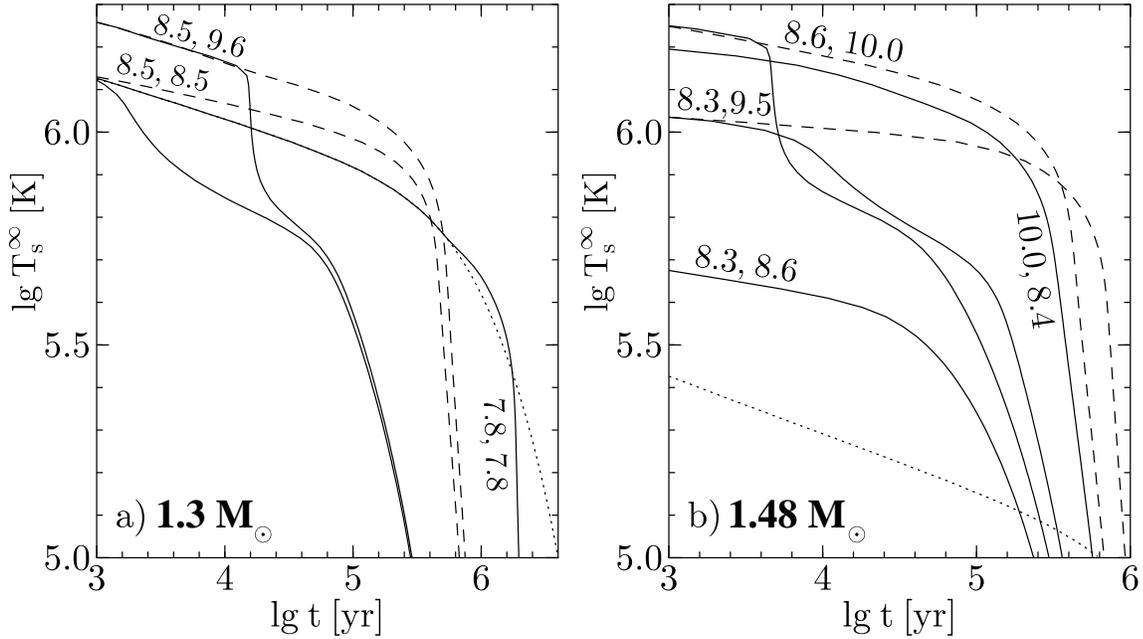}
\end{center}
\caption[]{
    Redshifted surface temperature $T_s^\infty$ versus age $t$
    for the standard (a) and enhanced (b) cooling
    of the 1.3\,$M_\odot$ and 1.48\,$M_\odot$ neutron stars,
    respectively.
    Dotted curves are for non--superfluid stars.
    Solid and dashed curves are for superfluid stars
    (marked with  ($\lg T_{cn}$, $\lg T_{cp}$))
    including and neglecting Cooper-pair neutrinos, respectively.
    Solid and dashed curves (7.8,7.8) in Fig.\ (a),  and
    (8.3,8.6) and (10,8.4) in Fig.\ (b) coincide.
    }
\label{fig4_1}
\end{figure*}

Figures \ref{qq_reca}a and b display the domains of
$T_{cn}$ and $T_{cp}$, where different neutrino mechanisms are
dominant. Figure \ref{qq_reca}a refers to the
standard cooling at $\rho \! = \! 2\, \rho_0$ and
$T= 3 \times 10^8$~K.
Figure \ref{qq_reca}b corresponds to the rapid cooling
at $ \rho \!= \! 5\, \rho_0$ and $T = 10^8$~K.
Dashes show the lines of $T=T_{cn}$ and $T=T_{cp}$,
which separate $T_{cn}$-$T_{cp}$ planes into four regions.
In the left down corners enclosed by these lines,
matter is nonsuperfluid. The right down corners
correspond to superfluidity of $n$ alone;
the left upper corners to superfluidity of $p$ alone,
and the right upper corners to joint $n$ and $p$ superfluidity.
One can observe a variety of dominant cooling mechanisms regulated
by the superfluidity. If both $n$ and $p$ superfluids are
very strong, they
switch off all the neutrino emission mechanisms involving
nucleons (and discussed here). In this regime,
a slow neutrino emission due to $ee$--bremsstrahlung
(Kaminker et al.\ \cite{kyh97}) survives and dominates. This
mechanism is rather insensitive to the superfluid state of dense matter.

\section{Models of cooling neutron stars} 
To illustrate the results of
Sects.\ 2 and 3 we have performed
simulations of neutron--star cooling.
We have used the same cooling code as described by
Levenfish \& Yakovlev (\cite{ly96}). The general relativistic
effects are included explicitly. The stellar cores are
assumed to have the same
equation of state (Prakash et al., \cite{pal88})
as has been used in Sect.\ 3.
The maximum neutron--star mass, for this equation of state,
is $1.73 \, M_\odot$.
We consider the stellar models
with two masses. In the first case, the mass is
$M=1.3 \, M_\odot$, the radius $R=11.87$~km, and
the central density
$\rho_c \! = \! 1.07 \! \times \! 10^{15}$ g cm$^{-3}$
is below the threshold ($\rho_{cr}$) of the direct Urca process;
this is an example of the standard cooling.
In the second case $M=1.48 \, M_\odot$,
$R=11.44$~km, and
$\rho_c \! = \! 1.376 \! \times \! 10^{15}$ g cm$^{-3}$.
The powerful direct Urca process
is open in a small central stellar kernel
of radius 2.32 km and mass $0.035 \, M_\odot$,
producing enhanced cooling.
Notice, that in calculations of the equation
of state of the stellar core in our earlier articles
(Levenfish \& Yakovlev \cite{ly96}, and references therein)
the parameter $n_0$ (standard saturation number density of baryons) was
set equal to 0.165 fm$^{-3}$. In the present article,
we adopt a more natural choice $n_0=0.16$ fm$^{-3}$
and use the model of the rapidly cooling star
with somewhat higher mass than before ($1.44\, M_\odot$).
The nucleon effective masses are set equal to 0.7
of their bare masses.

%
\begin{figure}[ht]                          
\begin{center}
\leavevmode
\epsfysize=8.5cm
\epsfbox{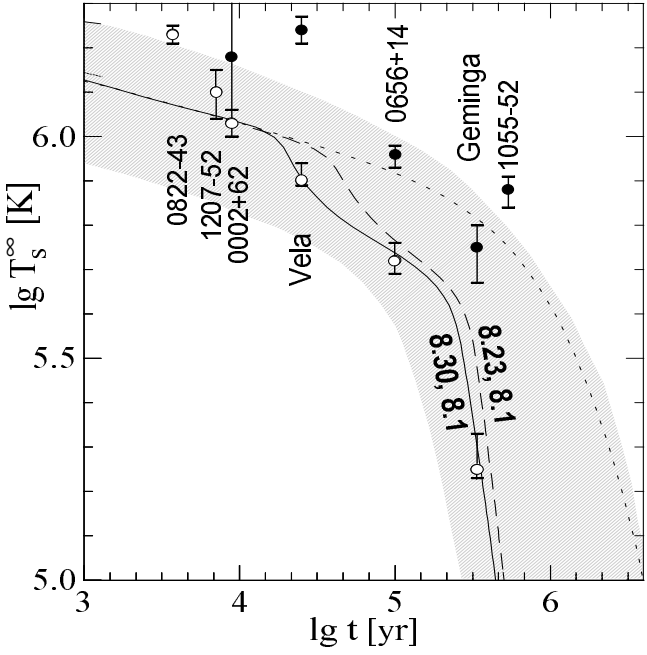}
\end{center}
\caption[]{
     Standard cooling curves (1.3\, $M_\odot$)
     compared with observations (Table~\ref{tab:NS_date}).
     Error bars are 90\%--95\% estimates of
     $T_s^\infty$ obtained by fitting the
     observed radiation spectra with black-body
     spectrum (filled circles) and atmosphere models (open circles).
     Dotted curve corresponds to a non-superfluid star.
     Solid and dashed curves labelled as
     in Fig.\ \protect{\ref{fig4_1}} are for superlfuid stars.
     Dashed region is formed by standard the curves
     for different $T_{cn}$ and $T_{cp}$.
     }
\label{fig4_2}
\end{figure}

For simplicity, the nucleons are assumed to be superfluid
everywhere in the stellar core. We suppose that
the proton superfluidity is of type (A),
while the neutron superfluidity is of type (B).
We make the simplified assumption that
$T_{cn}$ and $T_{cp}$
are constant over the stellar core and can be treated
as free parameters (see Sect.\ 3).

Our cooling code includes the main traditional neutrino
reactions in the neutron star core
(\ref{Murca})--(\ref{Durca}),
suppressed properly by neutron and proton superfluids
(Levenfish \& Yakovlev \cite{ly96}), supplemented by
the Cooper neutrino emission by neutrons and protons
(Sects.\ 2 and 3).
In addition we include the neutrino emission
due to electron-ion bremsstrahlung
in the neutron star crust
using an approximate formula by Maxwell (\cite{m79}).
The neutron--star heat capacity is assumed to
be the sum of the capacities of $n$, $p$, and $e$
in the stellar core affected by $n$ and $p$ superfluids
(Levenfish \& Yakovlev \cite{ly94b}).
We have neglected the heat capacity
of the crust due to small crustal masses for the chosen
stellar models.
Our cooling code is based on approximation of
isothermal stellar interior valid for a star of age
$t > (10$--$10^3)$~yr, inside which the thermal relaxation
is over. We use the relationship between
the surface and interior stellar temperature
obtained by Potekhin et al.\ (\cite{pcy97}). We assume that
the stellar atmosphere may contain light elements.
Then we can compare our results with observations of thermal radiation
interpreted using the hydrogen or helium atmosphere
models. However the mass of light elements
is postulated to be insufficient ($\la 10^{-14} M_\odot$)
to affect the cooling.

Figures \ref{fig4_1}a and b show typical cooling curves
(dependence of the effective surface temperature
$T_s^\infty$
versus stellar age $t$) for the neutron stars with superfluid cores.
The effective temperature is redshifted (as detected by a distant observer).
Figure \ref{fig4_1}a displays the standard cooling,
while \ref{fig4_1}b shows the enhanced cooling.
It is seen that Cooper-pair neutrino emission can be
either important or unimportant for the cooling of both types
depending on $T_{cn}$, $T_{cp}$ and $t$.
Its effect is to accelerate the cooling, and the stronger effect
takes place if
the neutron or proton superfluidity is
switched on in the neutrino cooling era ($t \la 10^5$ yr).
As mentioned in Sect.\ 3,
the neutrinos provided by
neutron pairing can dominate the neutrino emission
from the modified Urca process
at temperatures $T \la T_{cn} \la  10^9$~K. Therefore, if the neutron
superfluidity with $T_{cn} \la 10^9$ K is switched on
in the stellar core, the standard cooling is accelerated.
The acceleration is especially dramatic if the modified
Urca was suppressed by the proton superfluidity
before the onset of the neutron superfluidity ($T_{cn} \ll T_{cp}$).
Then the slow cooling looks like the enhanced one
(cf Figs.\ \ref{fig4_1}a and b).
On the other hand, Cooper-pair
neutrinos can be unimportant. The example is given
by the curve ($\lg T_{cn}=7.8$, $\lg T_{cp}=7.8$)
in Fig.\ \ref{fig4_1}a.
In this case, the star
enters the photon cooling era (with the photon surface
luminosity much larger than the neutrino one)
with the internal temperature $T \la 10^8$ K.
The superfluidity appears later,
and has naturally almost no effect on the cooling.
If $T_{cn} \ll T_{cp}$ the Cooper neutrino emission
by neutrons can dominate even the powerful direct
Urca process and accelerate the enhanced cooling
(Fig.\ \ref{fig4_1}b).
In other cases (e.g., the curves
($\lg T_{cn}=8.3$, $\lg T_{cp}=8.6$) and
($\lg T_{cn}=10$, $\lg T_{cp}=8.4$)) the effect of Cooper--pair
neutrinos on the enhanced cooling can be unimportant.

\newcommand{\rl}{\rule{0em}{1.8ex}}
\newcommand{\hhh}{\rule{1.4em}{0ex}}
\newcommand{\hhb}{\rule{1.2em}{0ex}}
\newcommand{\hhl}{\rule{2.4em}{0ex}}

\begin{table*}[pht]   
\caption{Observational data}
\label{tab:NS_date}
\begin{tabular}{|p{2.2cm}|p{1.5cm}|@{}p{12.3cm}|}
\hline                
\hline
 Source
    &
 $\;\;\lg t\rl,$ [yr]
    & \begin{tabular}[c]{@{$\;\;\;$}p{3.2cm}|p{1.9cm}|p{2.6cm}|p{3.25cm}@{}}
 Atmosphere model$^{\,b)}$
    &
 $\;\;\lg T_{\!s}^{\infty},\;$ [K]\rl
    &
  Confidence level$^{\,a)}$
    &
 Reference \rule{0em}{2.5ex}
                  \end{tabular} \\
\hline
\hline
RX$\,$J0822-43 & \hhh 3.57 &
  \begin{tabular}[c]{@{$\;\;\;$}p{3.2cm}|p{1.9cm}|p{2.6cm}|p{3.25cm}@{}}
   Hydrogen atmosphere &\hhb $6.23^{+0.02\rl}_{-0.02} $
                       &\hhl 95.5\%
                       & Zavlin et al.\ (\cite{zpt98b}) \\[0.5ex]
  \hline
            Black body & $\hhb 6.61^{+0.05\rl}_{-0.05}  $
                       &\hhl 95.5\%
                       & Zavlin et al.\ (\cite{zpt98b}) \\[0.5ex]
  \end{tabular}    \\
  \hline
        1E$\,$1207-52  &\hhh 3.85 &
  \begin{tabular}[c]{@{$\;\;\;$}p{3.2cm}|p{1.9cm}|p{2.6cm}|p{3.25cm}@{}}
   Hydrogen atmosphere &\hhb $6.10^{+0.05\rl}_{-0.06} $
                       &\hhl $\,$90\%
                       & Zavlin et al.\ (\cite{zpt98a}) \\[0.5ex]
  \hline
            Black body &\hhb $6.49^{+0.02\rl}_{-0.01}  $
                       &\hhl $\,$90\%
                       & Zavlin et al.\ (\cite{zpt98a}) \\[0.5ex]
  \end{tabular}    \\
  \hline
        RX$\,$J0002+62 & \hhh $3.95^{c)}$ &
  \begin{tabular}[c]{@{$\;\;\;$}p{3.2cm}|p{1.9cm}|p{2.6cm}|p{3.25cm}@{}}
   Hydrogen atmosphere &\hhb  $6.03^{+0.03\rl}_{-0.03} $
                       &\hhl 95.5\%
                       & Zavlin et al.\ (\cite{zpt98b}) \\[0.5ex]
  \hline
   Black body &\hhb $6.18^{+0.18\rl}_{-0.18} $
                       &\hhl 95.5\%
                       & Zavlin et al.\ (\cite{zpt98b}) \\[0.5ex]
  \end{tabular}    \\
  \hline
  \begin{tabular}{@{}l}
        PSR~0833-45 \\
        (Vela)
  \end{tabular}
                       & \hhh $4.4^{d)}$ &
  \begin{tabular}[c]{@{$\;\;\;$}p{3.2cm}|p{1.9cm}|p{2.6cm}|p{3.25cm}@{}}
   Hydrogen atmosphere &\hhb $5.90^{+0.04\rl}_{-0.01}$
                       &\hhl $\,$90\%
                       & Page et al.\ (\cite{psz96}) \\[0.5ex]
  \hline
   Black body &\hhb $6.24^{+0.03\rl}_{-0.03} $
                       &\hhl $\;\;\;$---
                       & \"{O}gelman (\cite{o95})\\[0.5ex]
  \end{tabular}    \\
  \hline
   PSR~0656+14 & \hhh 5.00 &
  \begin{tabular}[c]{@{$\;\;\;$}p{3.2cm}|p{1.9cm}|p{2.6cm}|p{3.25cm}@{}}
   Hydrogen atmosphere &\hhb $5.72^{+0.04\rl}_{-0.02} $
                       &\hhl $\;\;\;$---
                       & Anderson et al.\ (\cite{acprt93}) \\[0.5ex]
  \hline
   Black body &\hhb $5.96^{+0.02\rl}_{-0.03} $
                       &\hhl $\,$90\%
                       & Possenti et al.\ (\cite{pmc96}) \\[0.5ex]
  \end{tabular}    \\
  \hline
  \begin{tabular}{@{}l}
     PSR~0630+178\\
     (Geminga)
  \end{tabular}
                       & \hhh 5.53 &
  \begin{tabular}[c]{@{$\;\;\;$}p{3.2cm}|p{1.9cm}|p{2.6cm}|p{3.25cm}@{}}
   Hydrogen atmosphere &\hhb  $5.25^{+0.08\rl}_{-0.01} $
                       &\hhl $\,$90\%
                       & Meyer et al.\ (\cite{mpm94}) \\[0.5ex]
  \hline
   Black body &\hhb $5.75^{+0.05\rl}_{-0.08} $
                       &\hhl $\,$90\%
                       & Halpern, Wang (\cite{hw97}) \\[0.5ex]
  \end{tabular}    \\
  \hline
  PSR~1055-52 & \hhh 5.73 &
  \begin{tabular}[c]{@{$\;\;\;$}p{3.2cm}|p{1.9cm}|p{2.6cm}|p{3.25cm}@{}}
  Black body &\hhb $5.88^{+0.03\rl}_{-0.04} $
                       &\hhl $\;\;\;$---
                       & \"{O}gelman (\cite{o95}) \\[0.5ex]
  \end{tabular} \\
  \hline
  \multicolumn{3}{@{}l@{}}{
  \begin{tabular}{@{}l@{}}
  \rule{0em}{3ex}$^{a)}\,\rl${\footnotesize
        Confidence level of $T_s^\infty$ (90\% and 95.5\% correspond
        to $1.64\sigma$ and $2\sigma$, respectively);
        dash means that the level is not}\\[-0.7ex]
  $\phantom{^{ a) }}\,\rl${\footnotesize
        indicated in cited references.
        }\\[-0.7ex]
    $^{b)}\,$\rl{\footnotesize
         Method for interpretation of observation.
         }\\[-0.7ex]
  $^{c)}\,$\rl{\footnotesize
       The mean age taken according to Craig et al.\ (\cite{chp97}).
       }\\[-0.7ex]
  $^{d)}\,$\rl{\footnotesize According to Lyne et al.\ (\cite{lpgc96}). }
\end{tabular}
 }\\
\end{tabular}
\end{table*}

Figure \ref{fig4_2} compares the
standard cooling curves ($1.3 \, M_\odot$)
with
the available
observations of thermal radiation from
isolated neutron stars.
The observational data
are summarized in Table 2.
We include the data on four pulsars
(Vela, Geminga, PSR 0656+14, PSR 1055-52) and
three radioquiet objects (RX J0822-43, 1E 1207-52, RX J0002+62)
in supernova remnants.
The pulsar
ages are the dynamical ages except for Vela, where
new timing results by Lyne et al.\ (\cite{lpgc96}) are used.
Ages of radioquiet objects are associated to ages of their
supernova remnants.
The error bars give the confidence
intervals of the redshifted
effective surface temperatures obtained by two different methods.
The first method consists in fitting the observed
spectra by neutron--star hydrogen and/or helium
atmosphere models (open circles), and the second one
consists in fitting by the blackbody spectrum (filled circles).
Dashed region encloses all standard cooling curves
for a $1.30\,M_\odot$ star with different $T_{cn}$ and $T_{cp}$
(from $10^6$~K to $10^{10}$~K).
Notice that the Cooper-pair neutrinos
introduce into the cooling theory some new ``degree of freedom''
which helps one to fit the observational data.
For instance, we present a solid curve which
hits five error bars for the atmosphere models at once.
We would not be able to find a similar cooling
curve if we neglected the Cooper-pair neutrino emission.
The dashed curve in Fig.\ \ref{fig4_2}
shows that Cooper-pair neutrinos
make the standard cooling
very sensitive to the superfluid parameters.
A minor change even of the one superfluid parameter yields
quite a different cooling curve, which hits two error bars only.
Such a sensitivity is important for constraining $T_{cn}$ and $T_{cp}$ from
observations. 

Thus, the majority of observations of
thermal radiation from isolated neutron stars can be interpreted
by the standard cooling with quite a moderate nucleon
superfluidity in the stellar core. These moderate
critical temperatures do not contradict to a wealth of
microscopic calculations of $T_{cp}$ and $T_{cn}$.
Notice that it is easier for us to explain
the ``atmospheric'' surface temperatures than
the blackbody ones.
This statement can be considered as
an indirect argument in favour of the atmospheric
interpretation of the thermal radiation.
Although the theory of neutron--star atmospheres
is not yet complete (e.g., Pavlov \& Zavlin \cite{pz98}) the
atmospheric fits give more reasonable
neutron--star parameters (radii, magnetic fields, distances, etc.)
which are in better agreement with the parameters obtained
by other independent methods.

Even in our simplified model (one equation
of state, two fixed neutron--star masses,
constant $T_{cn}$ and $T_{cp}$ over the stellar core)
it is possible to choose quite definite superfluid
parameters to explain most of observations.
However, one needs more elaborated models
of cooling neutron stars to obtain reliable information
on superfluid state of the neutron star cores.
We expect to develop such models in the future
making use of the results obtained in the present article.

\section{Conclusions} 
We have reconsidered the neutrino emission rate due to
Cooper pairing of nucleons in the neutron star cores (Sect.\ 2).
We have presented the results in the form convenient
for practical implications in three cases: (A) singlet--state $^1$S$_0$
pairing, (B) triplet--state $^3$P$_2$ pairing with zero
projection of the Cooper pair momentum ($m_J=0$) onto quantization
axis, and (C) triplet--state pairing with maximum ($m_J=2$)
momentum projection. For the singlet-state
pairing of neutrons, our results
agree with those by Flowers et al.\ (\cite{frs76}).
For the triplet-state
pairing our consideration is original. Notice an essential
difference of the Cooper-pair neutrino emissivities for singlet and triplet-state superfluids and also for neutrons and protons.
In Sect.\ 3 we have analysed the efficiency of
the Cooper-pair neutrino emission at different densities in the
neutron star cores as compared with the traditional neutrino
production mechanisms including a powerful direct
Urca process allowed at high densities.
Contrary to the non-superfluid cores where the
main neutrino emission is produced either by the modified
or by the direct Urca processes (depending on equation of state
and density), very different neutrino mechanisms can
dominate in the superfluid cores at certain temperatures $T$ and
superfluid critical temperatures $T_{cn}$ and $T_{cp}$.
In particular, neutrinos produced by pairing of
neutrons can be very important if $T \la T_{cn} \ll T_{cp}$.
The importance of these neutrinos in the standard and
rapid cooling of the neutron stars has been analysed
in Sect.\ 4. We show that, under certain conditions,
neutrinos provided by pairing of neutrons
can greatly accelerate both standard and enhanced cooling
of middle-age neutron stars ($t$= $10^4$--$10^5$ yr).
In particular, the accelerated standard cooling
can mimic rapid cooling of the stars.
The Cooper-pair neutrinos modify the cooling curves
and enable us to explain observations of thermal
radiation of several neutron stars by one cooling
curve at once. This confirms the
potential ability (Page \& Applegate, \cite{pa92})
to constrain the fundamental parameters of superdense matter
in the neutron star cores, the critical temperatures of neutron
and proton superfluids,
by comparing theory and observation of
neutron stars.

\begin{acknowledgements} We are grateful to D.A.\ Baiko,
P.\ Haensel, C.J.\ Pethick,
Yu.A.\ Shibanov, D.A.\ Varshalovich and D.N.\ Voskresensky
for useful discussions.
This work was supported in part by RFBR (grant 96-02-16870a),
RFBR-DFG (grant 96-02-00177G)
and INTAS (grant 96-0542).
\end{acknowledgements}


\begin{thebibliography}{}

\bibitem[1985]{ao85}
Amundsen L., {\O}stgaard E., 1985, Nucl.\ Phys.\ A442, 163

\bibitem[1993]{acprt93}
Anderson S., C\'{o}rdova F., Pavlov G.G., Robinson C.R.,  Thompson R.J.,
1993, ApJ 414, 867

\bibitem[1998]{bb98}
Balberg S., Barnea N., 1998, Phys.\ Rev.\ 57C, 409

\bibitem[1992]{bcll92}
Baldo M., Cugnon J., Lejeune A., Lombardo U., 1992, Nucl.\ Phys.\ A536, 349

\bibitem[1997]{chp97}
Craig W.W., Hailey Ch.J., Pisarski R.L., 1997,
ApJ 488, 307

\bibitem[1976]{frs76}
Flowers E., Ruderman M., Sutherland P., 1976, ApJ 205, 541

\bibitem[1979]{fm79}
Friman B.L., Maxwell O.V., 1979, ApJ 232, 541



\bibitem[1997]{hw97}
Halpern J., Wang F., 1997, ApJ 477, 905

\bibitem[1997]{kyh97}
Kaminker A.D., Yakovlev D.G., Haensel P., 1997, A\&A 325, 391

\bibitem[1991]{lpph91}
Lattimer J.M., Pethick C.J., Prakash M., Haensel P.,
1991, Phys.\ Rev.\ Lett. 66, 2701

\bibitem[1994a]{ly94a}
Levenfish K.P., Yakovlev D.G., 1994a, Astron.\ Lett. 20, 43

\bibitem[1994b]{ly94b}
Levenfish K.P., Yakovlev D.G., 1994b, Astron.\ Rep. 38, 247

\bibitem[1996]{ly96}
Levenfish K.P., Yakovlev D.G., 1996, Astron.\ Lett. 22, 56

\bibitem[1998]{lsy98}
Levenfish K.P., Shibanov Yu.A., Yakovlev D.G., 1998,
Physica Scripta T77, 79

\bibitem[1980]{lp80}
Lifshitz E.M., Pitaevskii L.P., 1980, Statistical Physics,
part 2, Pergamon, Oxford

\bibitem[1996]{lpgc96}
Lyne A.G., Pritchard R.S., Graham-Smith F., Camilo F., 1996,
Nat.\ 381, 497

\bibitem[1979]{m79}
Maxwell O.V., 1979, ApJ 231, 201

\bibitem[1994]{mpm94}
Meyer R.D., Pavlov G.G., M\'{e}sz\`{a}ros P., 1994, ApJ 433, 265

\bibitem[1995]{o95}
\"{O}gelman H., 1995.\ 
In: Alpar M.A., Kizilo\u{g}lu \"{U}, van Paradijs J.\ (eds.) 
Lives of Neutron Stars, 
Kluwer Academic Publishers, Dordrecht, p.\ 101

\bibitem[1990]{o90}
Okun' L.B., 1990, Leptons and Quarks, Nauka, Moscow (in Russian)

\bibitem[1998]{p98}
Page D., 1998.
In:
Buccheri R., van Paradijs J., Alpar M.A.\ (eds.)
The Many Faces of Neutron Stars, Kluwer, Dordrecht, p.\ 538


\bibitem[1992]{pa92}
Page D., Applegate J.H., 1992, ApJ 394, L17

\bibitem[1996]{psz96}
Page D., Shibanov Yu., Zavlin V., 1996.\ In:
Zimmermann H.U., Tr\"{u}mper J.E., Yorke H.\ (eds.)
R\"{o}ntgenstrahlung from the Universe,
MPE, Garching, p.\ 173

\bibitem[1998]{pz98}
Pavlov G.G., Zavlin V.E., 1998.\
In: 
Shibazaki N., Kawai N., Shibata S., Kifune T.\ (eds.)
Neutron Stars and Pulsars,
Universal Academy Press, Tokyo, p.\ 327

\bibitem[1992]{p92}
Pethick C.J., 1992, Rev.\ Mod.\ Phys. 64, 1133

\bibitem[1997]{pcy97}
Potekhin A.Yu., Chabrier G., Yakovlev D.G., 1997, A\&A 323, 415

\bibitem[1988]{pal88}
Prakash M., Ainsworth T.L., Lattimer J.M., 1988,
Phys.\ Rev.\ Lett. 61, 2518

\bibitem[1996]{pmc96}
Possenti A., Mereghetti S., Colpi M., 1996, A\&A 313, 565

\bibitem[1997]{svsww97}
Schaab Ch., Voskresensky D., Sedrakian A.D., Weber F.,
Weigel M.K., 1997, A\&A 321, 591

\bibitem[1998]{sbs98}
Schaab Ch., Balberg S., Schaffner-Bielich J., 1998,
ApJL 504, L99

\bibitem[1993]{tt93}
Takatsuka T., Tamagaki R., 1993, Prog.\ Theor.\ Phys.\  112, 27

\bibitem[1970]{tt97}
Takatsuka T., Tamagaki R., 1997, Prog.\ Theor.\ Phys.\  97, 345

\bibitem[1970]{t70}
Tamagaki R., 1970, Prog.\ Theor.\ Phys. 44, 905

\bibitem[1986]{vs86}
Voskresensky D., Senatorov A., 1986, Sov.\ Phys.--JETP 63, 885

\bibitem[1987]{vs87}
Voskresensky D., Senatorov A., 1987, Sov.\ J.\ Nucl.\ Phys. 45, 411

\bibitem[1995]{yl95}
Yakovlev D.G., Levenfish K.P., 1995, A\&A 297, 541

\bibitem[1998]{ykl98}
Yakovlev D.G., Kaminker A.D., Levenfish K.P., 1998.\
In: 
Shibazaki N., Kawai N., Shibata S., Kifune T.\ (eds.)
Neutron Stars and Pulsars,
Universal Academy Press, Tokyo, p.\ 195

\bibitem[1998a]{zpt98a}
Zavlin V., Pavlov G.G., Tr\"{u}mper J., 1998a, A\&A 331, 821

\bibitem[1998b]{zpt98b}
Zavlin V., Pavlov G.G., Tr\"{u}mper J., 1998b,
X---ray emission from neutron stars in two supernova remnants,
A\&A (accepted)

\end{thebibliography}
\end{document}